\documentclass[journal=apchd5,manuscript=article]{achemso}
\usepackage[version=3]{mhchem}
\setkeys{acs}{doi=true}
\newcommand{\doi}[1]{\href{https://doi.org/#1}{\nolinkurl{#1}}}

\usepackage{graphicx} 
\usepackage{amsmath}

\usepackage{amssymb}
\usepackage{siunitx}
\usepackage{physics}

\usepackage{hyperref}
\hypersetup{pdfstartview={FitH},pdfpagemode={UseNone},
            colorlinks,linkcolor=blue, citecolor=blue, urlcolor=blue,
            bookmarksopen=true, pdfnewwindow=true}
\usepackage[all]{hypcap}
\graphicspath{{Figures}{}}

\title{Angle multifunctional dichroism in metasurfaces}
\author{Neuton Li}
\email{Neuton.Li@anu.edu.au}
\affiliation{ARC Centre of Excellence for Transformative Meta-Optical Systems (TMOS), Department of Electronic Materials Engineering, Research School of Physics, The Australian National University, Canberra, ACT~2600, Australia}

\author{Jihua Zhang} 
\affiliation{Songshan Lake Materials Laboratory, Dongguan, Guangdong 523808, P. R. China}
\alsoaffiliation{ARC Centre of Excellence for Transformative Meta-Optical Systems (TMOS), Department of Electronic Materials Engineering, Research School of Physics, The Australian National University, Canberra, ACT~2600, Australia}

\author{Dragomir N. Neshev}
\affiliation{ARC Centre of Excellence for Transformative Meta-Optical Systems (TMOS), Department of Electronic Materials Engineering, Research School of Physics, The Australian National University, Canberra, ACT~2600, Australia}

\author{Andrey A. Sukhorukov}
\affiliation{ARC Centre of Excellence for Transformative Meta-Optical Systems (TMOS), Department of Electronic Materials Engineering, Research School of Physics, The Australian National University, Canberra, ACT~2600, Australia}

\begin{document}

\begin{abstract}
We demonstrate metasurfaces with strong polarization dichroism that depends on the angle of incidence. We present original designs obtained through topology optimization that selectively transmit specific linear or circular polarizations at different incident angles, while the orthogonal polarization transmission is suppressed. The designed metasurfaces exceed 95\% transmission efficiency and $50\times$ extinction ratio within the target angle ranges. The experimental characterization of fabricated metasurfaces confirms the desired operation with 90\% transmission efficiency and $10\times$ extinction ratio. These results provide important insights to non-local and $k$-space engineering of metasurface response and results reveal new opportunities for future multi-functional and angle-selective polarization devices that can find applications in specialized optical instruments and end-user devices.
\end{abstract}


\section{Introduction}
Subwavelength scale material structuring has enabled researchers to engineer metasurfaces to scatter light in complex and tailored ways. Consequently, there is a strong interest in the community to understand and explore the furthest extent of light shaping with metasurfaces. One particular but important aspect is to investigate the effects of varying incidence angles of light on a metasurface. The well-known Fresnel equations describe the effect between varying oblique incidence, and the transmittance and reflectance of different polarizations for unpatterned surfaces. Gratings and metasurfaces alter the polarization behaviour to different degrees, and expand the possibilities for novel functionalities. By fully controlling the polarization behaviour of metasurfaces under different incidence conditions, we may not only combine functionalities onto a single device, it becomes possible to shape the wavefront of electromagnetic fields non-locally. Such ideas may develop and translate into applications that include analogue optical processing \cite{Wan2021ADetection,Zhou2020AnalogMetasurfaces, Kwon2018NonlocalProcessing,He2022ComputingReview} and wavefront manipulation \cite{Song2021Non-localTracking,Overvig2022DiffractiveMetasurfaces,Malek2022MultifunctionalMetasurfaces}.  

There have been numerous demonstrations of metasurfaces with polarizing characteristics, including transmissive polarizers \cite{Rubin2021PolarizationMetasurfaces,Zaidi2021GeneralizedMetasurfaces, Dorrah2021MetasurfacePath,Wang2021ArbitraryPolarizers,Wang2023Metasurface-BasedPolarizer}, polarization beam splitters \cite{Guo2017DielectricSplitters,Khorasaninejad2015EfficientMetasurface,Li:2024-66002:ADPN} and polarization holograms \cite{Zhao2022ControllableMetasurface,A2021JonesMetasurfaces, Deng2018DiatomicHolography, Zaidi2024Metasurface-enabledImaging}. More recently, multiplexing different polarization behaviours with changing incidence at oblique angles was explored. One demonstration showed continuous tuning of birefringence with an incident angle on a dielectric metasurface \cite{Zhujun2021ContinuousConversion}. In this way, the metasurface could transform input polarizations into a different output state that is dependent on the incident angle, however there was no dichroism as all polarizations were transmitted with the same efficiency. Another study demonstrated linear and circular metasurface dichroism that is dependent on the incident azimuth angles \cite{Huang2020Dual-FunctionalDichroism}, however it operated in reflection and relied on metal-insulator-metal (MIM) geometry that places restrictions on integration in optical devices. Yet, it remained an outstanding challenge to realize angular-dependent dichroism and arbitrary polarization bases in transmission through all-dielectric metasurfaces.

\begin{figure}[!t]
    \centering
    \includegraphics[width = 0.75\linewidth]{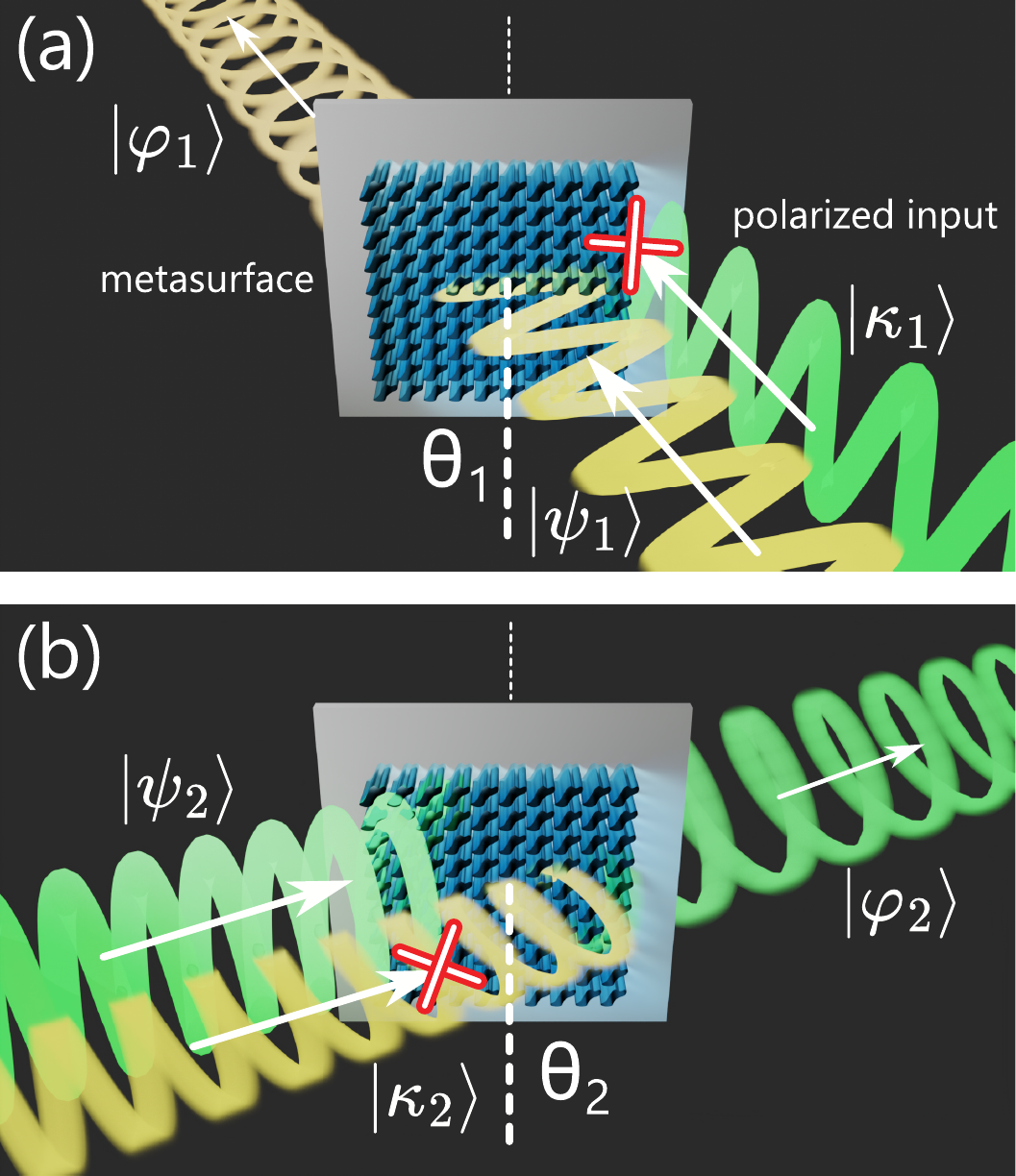}
    \caption{(a) A metasurface that exhibits different dichroism depending on the angle of incidence. Transmission of one polarization $\ket{\psi_1}$ is maximal for angle of incidence at $\theta_1$, while the transmission of the orthogonal polarization $\ket{\kappa_1}$ is either blocked or reflected. (b) At a different incident angle $\theta_2$, the metasurface exhibits dichroism for another set of polarizations.}
    \label{fig:schem_1}
\end{figure}

The difficulty in engineering the desired polarization response at different angles of incidence lies in the multi-objective functionality of the metasurface, and additionally fulfilling all the necessary properties in a single-layer metasurface would be most attractive.
Many researchers have employed inverse design with free-form topology \cite{Molesky2018InverseNanophotonics,Li2022EmpoweringApplications} as an approach to tackle the designing of multi-functional metasurfaces for various applications. The optimized metasurface designs may have complicated non-intuitive geometries, which most remarkably can achieve the desired electromagnetic scattering responses. 
In this work, we employ the inverse design approach and successfully obtain new dielectric metasurface structures that feature distinct types of polarization dichroism in transmission at varying oblique angles of incidence, overcoming the limitations of the previously considered structures.  Furthermore, we fabricate the metasurfaces and experimentally demonstrate their desired operation.

\section{General framework of metasurface design}

The scattering from a metasurface is described with a Jones matrix that connects the complex amplitude of the incident and transmitted fields 
\begin{equation}
    \begin{pmatrix}
        \varphi_H \\ \varphi_V
    \end{pmatrix} = 
    \begin{pmatrix}
        t_{HH} & t_{HV} \\ t_{VH} & t_{VV} 
    \end{pmatrix}
    \begin{pmatrix}
        \psi_H \\ \psi_V
    \end{pmatrix} ,
\end{equation}
which we can write in an equivalent compact form as $\ket{\varphi} = J\ket{\psi}$. This  expresses the scattering as a transformation, via $J$, of the incident Jones vector $\ket{\psi}$ into the output state $\ket{\varphi}$, in the horizontal ($H$) and vertical ($V$) polarization basis.

There are certain conditions that metasurfaces must or very strongly obey. The first is optical reciprocity due to the fact that we only consider metasurfaces without gain, magnetic media nor nonlinearities~\cite{Asadchy:2020-1684:PIEEE}. The second condition is the restricion on the allowable forms of the Jones matrices that correspond to symmetries of the system, including that of the metasurface unit cell. If one enforces unitarity (i.e. conservation of transmitted power) alongside reciprocity and symmetry, then it is found that Jones matrices obey the following relationship: \cite{Zhujun2021ContinuousConversion}
\begin{equation}
\label{eq:J_unitary}
    J(\theta) \approx J(-\theta)^T \,.
\end{equation}
Using these principles, the researchers designed a device that has birefringence which can be tuned continuously with angle of incidence~\cite{Zhujun2021ContinuousConversion}. Birefringence is the property where different polarizations observe a different effective refractive index, and hence polarization amplitudes may be preserved and undergo a rotation at the same time.  

Importantly, the property of unitarity is not necessarily required in all optical components, as evidenced in polarizers. Dichroism occurs when different polarizations have different transmission efficiencies, so one may consider dichroism to be a generalization of birefreingence. We investigate dichroic regimes that violate Eq.~(\ref{eq:J_unitary}), which allows us to design metasurface devices that produce novel polarization phenomena in combination with varying obliques angles of incidence as sketched in Fig.~\ref{fig:schem_1}. In the general case, at an incidence angle of $\theta_1$, the input state $\ket{\psi_1}$ may be converted by the metasurface, and transmitted as $\ket{\varphi_1}$. However, the orthogonal state $\ket{\kappa_1}$, defined by $\bra{\kappa_1}\ket{\psi_1} = 0$, will have low transmission. Then as the angle of incidence changes to $\theta_2$, a different state $\ket{\psi_2}$ will be maximally transmitted to $\ket{\varphi_2}$, and its corresponding orthogonal state will be minimally transmitted $\ket{\kappa_2}$.

We optimize our metasurfaces with specific polarization behaviour using inverse-design, which is especially well suited for the current task because it allows for seemingly arbitrary geometries to be generated. {Our inverse-design approach is based on maximizing a figure of merit (FOM) via gradient descent, and is adapted from the Metanet codebase \cite{Jiang2020MetaNet:Research} that is coupled with the rigorous coupled-wave analysis, (RETICOLO software package \cite{Hugonin:2101.00901:ARXIV}) to model the electromagnetics.} With Lorentz reciprocity, it is possible to determine the material gradient for all positions in the unit cell with only one forward and one adjoint simulation for each set of illumination conditions~\cite{Molesky2018InverseNanophotonics, Li2022EmpoweringApplications, Jensen2011TopologyNano-photonics}. This allows extremely accurate and efficient determination of the material derivatives. Our gradient-based inverse design optimisation approach converges to the optimal structure within only 150 iterations. We also have incorporated binarization filters and impose minimum structural feature sizes through Gaussian blurring~\cite{Wang2019RobustMetasurfaces} to ensure fabricability.   

\section{Linear Polarization}

\begin{figure}[!t]
    \centering
    \includegraphics[width = 0.95\linewidth]{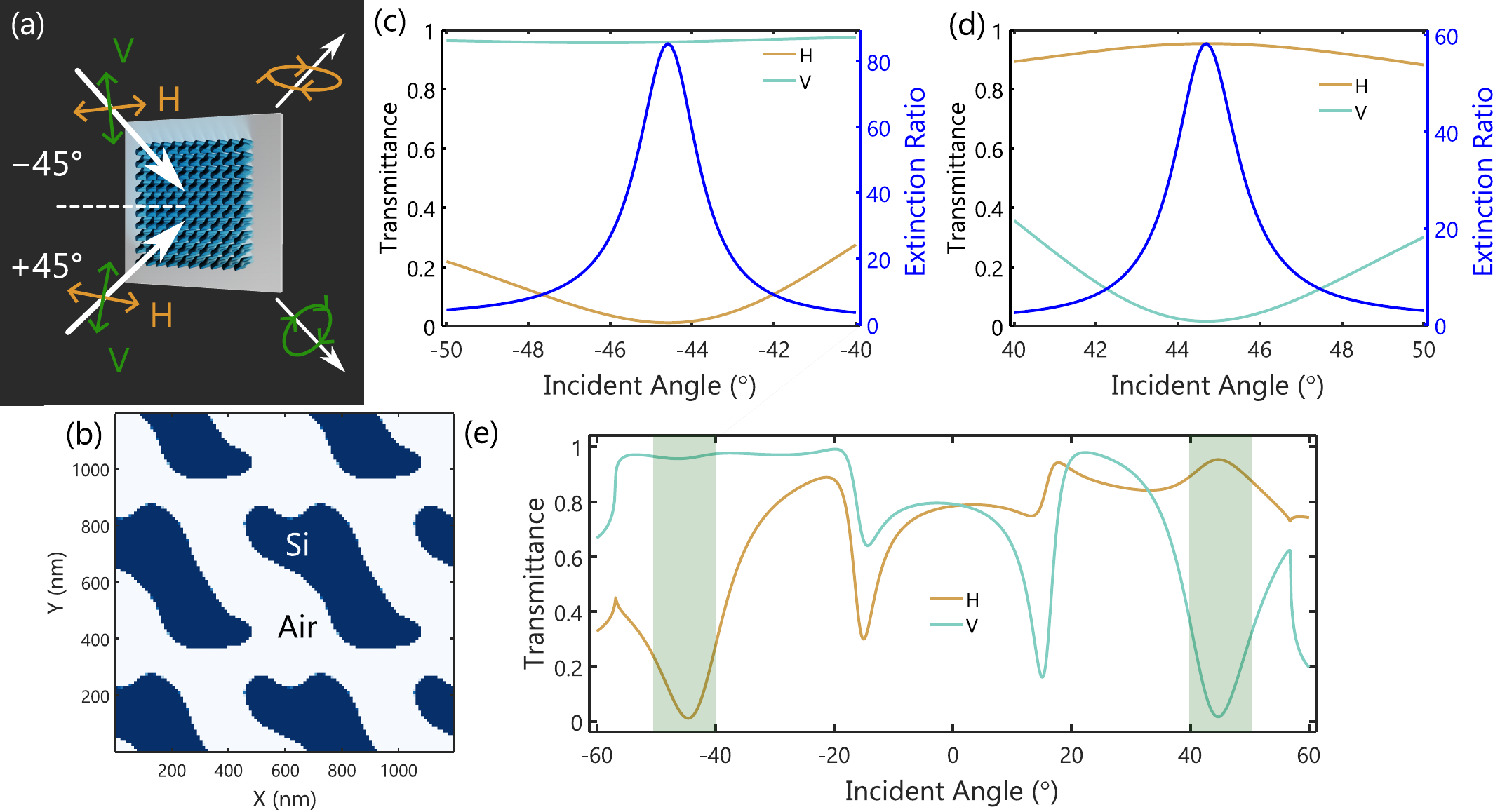}
    \caption{(a) Intended operating scheme for the metasurface. At $-\SI{45}{\degree}$ incidence, the metasurface only transmits V polarization, and at $+\SI{45}{\degree}$ incidence, it only transmits H polarization. (b) Final optimized design of the metasurface, with regions labelled for silicon and air. Transmitted efficiencies for H and V polarizations with their respective extinction ratios at (c) -\SI{45}{\degree} and (d) +\SI{45}{\degree} incidence angles. (e) Transmitted efficiencies of H and V polarizations over a range of oblique angular incidences, showing varying dichroism.   }
    \label{fig:lin_pol_sim}
\end{figure}

We first explore the design of a metasurface with linear polarizing functionality that is dependent on the incident angle. This dichroism is achieved by only transmitting certain polarizations at certain angles of incidence. In this example, our metasurface only transmits vertical polarization $\ket{V}$ at \SI{-45}{\degree} incident angle, and transmits only horizontal polarization $\ket{H}$ at $+\SI{45}{\degree}$, as sketched in  Fig.~\ref{fig:lin_pol_sim}(a). To find the metasurface design, we optimize for the performance by maximizing the figure of merit defined as
\begin{equation}
    FOM = \qty(\abs{J_{-\theta}\ket{V}}^2 - \abs{J_{-\theta}\ket{H}}^2) \qty(\abs{J_{+\theta}\ket{H}}^2 - \abs{J_{+\theta}\ket{V}}^2).
    \label{eq:FOM}
\end{equation}
The first term of the product favors the transmission of the polarization input state $\ket{V}$ over $\ket{H}$ at incident angle $-\theta$. The second term is for the angle $+\theta$ and is orthogonal to the first term. At the beginning of the optimization, the initial device is set to be a random distribution of intermediate refractive indices. Over iterations, the device gradually converges to an optimal material distribution, following the derivative of the FOM. The maximum value of the FOM defined in Eq.~(\ref{eq:FOM}) is 1, corresponding to unitary transmission of the intended polarization, and no transmission of the undesired polarization state at all. Our optimisation converges to a design where ${\rm FOM} = 0.94$, which indicates good performance close to the ideal case. 

We design metasurfaces that are comprised of \SI{1000}{\nano\metre} thick silicon patterns atop of sapphire substrate, and intended for operation at the $\lambda = \SI{1550}{\nano\metre}$ wavelength. The resulting optimized structure has a non-unintuitive geometry and does not have any clear geometric symmetries, see Fig.~\ref{fig:lin_pol_sim}(b). The metasurface exhibits distinct linear dichroism for $\SI{\pm45}{\degree}$ incidence. At $-\SI{45}{\degree}$ incidence  (Fig.~\ref{fig:lin_pol_sim}(c)), transmission of $\ket{V}$ is maximal while $\ket{H}$ is suppressed. For $+\SI{45}{\degree}$ incidence  (Fig.~\ref{fig:lin_pol_sim}(d)), the orthogonal case is true. At both target incident angles, the transmission efficiency is very high, approaching unity for the selected polarizations. We further observe that the maximum extinction ratios, defined as ${\rm ER} = {T_\parallel}/{T_\perp}$ of transmissions for orthogonal states at the target incident angles of $-\SI{45}{\degree}$ and $+\SI{45}{\degree}$, are approximately 80 and 60, respectively. The full width half maximum (FWHM) of the ER peak is approximately \SI{3}{\degree} around the target angles. We also show the metasurface transmission for the $\ket{H}$ and $\ket{V}$ inputs over a range of oblique angles of incidence, from \SI{-60}{\degree} to +\SI{60}{\degree}, see Fig.~\ref{fig:lin_pol_sim}(e). We observe that the intermediate oblique incident angles exhibit weak or no linear dichroism. 

We then perform singular value decomposition to characterise the input and output polarizations with maximal transmission. The two singular values provide the efficiency of transmission corresponding to the two right singular vector polarizations. We find that the largest input singular polarization is slightly elliptical, but is very close to $\ket{V}$ at \SI{45}{\degree}, and close to $\ket{H}$ at $+\SI{45}{\degree}$. The output states are not specified during the optimization because we do not require them to be any particular state. Thus, they generally will be an elliptical state. 

\begin{figure}[!t]
    \centering
    \includegraphics[width = 0.95\linewidth]{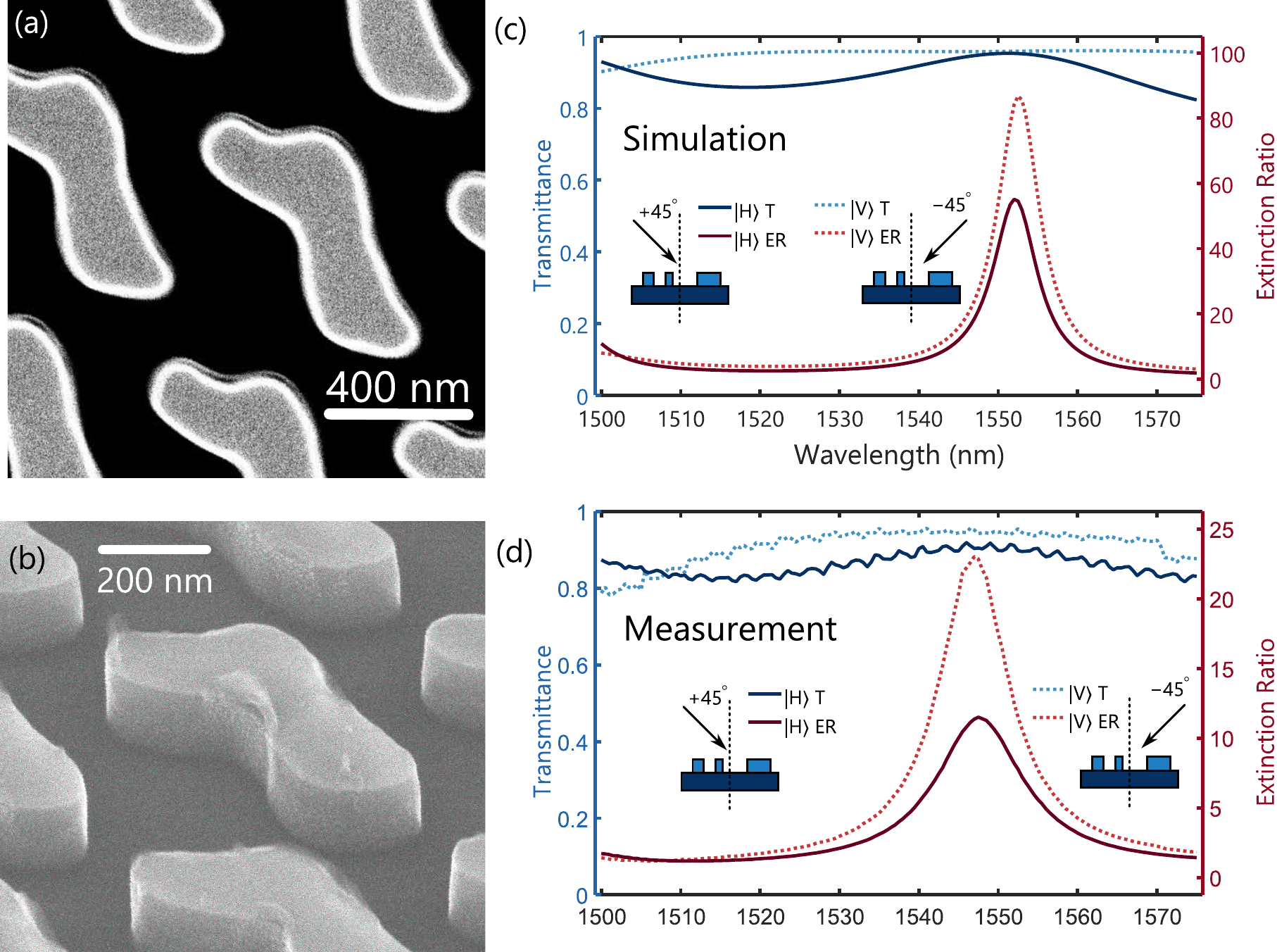}
    \caption{SEM image of fabricated metasurface design from (a) top down and (b) side angle view. The nano-structures are silicon. (c) Comparing simulated transmission efficiency to (d) measured values over a wavelength range of \SIrange{1500}{1575}{\nano\metre}. For both cases, $\ket{H}$ polarization is incident at $+\SI{45}{\degree}$, and $\ket{V}$ polarization is incident at $-\SI{45}{\degree}$  }
    \label{fig:lin_pol_exp}
\end{figure}

We then fabricate our devices with standard silicon processing. The pattern is written with electron beam lithography onto a polymer resist. Then an aluminium oxide mask is deposited for the final step of reactive ion etching. The final device has structures with high fidelity to the design (Figs.~\ref{fig:lin_pol_exp}(a,b)). From our simulations, the metasurface is expected to operate at $\lambda = \SI{1550}{\nano\metre}$ with a small bandwidth of operation (Fig.~\ref{fig:lin_pol_exp}(c)). In our measurements, we polarize our source into $\ket{V}$ and $\ket{H}$ and tilt our metasurface such that the oblique angle relative to the laser beam is $\SI{\pm45}{\degree}$. We measure from our metasurface a transmission efficiency of greater than 95\% for $\ket{V}$ at $-\SI{45}{\degree}$, and for $\ket{H}$ at $+\SI{45}{\degree}$ (Fig.~\ref{fig:lin_pol_exp}(d)). The peak ER measured is 20 for incident angle $-\SI{45}{\degree}$, and 10 for incident angle $+\SI{45}{\degree}$. We observe very similar transmission profiles between simulation and our measured results, indicating high quality fabrication.

\section{Circular Polarization}
To achieve circular dichroism in metasurfaces, the system must exhibit some form of chirality. Intrinsic chriality may be manifested through asymmetric 3D or planar nanostructures \cite{Shi2022PlanarContinuum, Chen2023ObservationContinuum,Wang2020GiantMetasurfaces,Zhu2018GiantNanostructures,Li2021LosslessWave}. In contrast, multiple studies also have shown that circular dichroism and circular birefringence can be achieved in a metasurface that has mirror symmetry in the plane of incidence; for example structures that exhibit $C_2$ symmetry \cite{Plum2008OpticalMetamaterial,Plum:2009-113902:PRL,Cao2016Dual-bandMetasurface,Lai2020AngleMetamaterial,Mao2020ExtrinsicallyRegion}. In this case, the incidence of light on the metasurface must be oblique. The system as a whole can be considered to be extrinsically chiral, even if the structures are themselves non-chiral. In this study, we use inverse-design to maximize the degree of circular dichroism and using the varying angle of incidence as a tuning mechanism.

We explore and investigate the degree of circular dichroism at oblique incidence in this scheme. We first use symmetry arguments to analyze the allowable forms of the Jones matrix at oblique incidence for symmetrical structures. Such a reasoning leads to our proposal of a metasurface with continuous angular tuning of its circular and linear dichroism. At oblique incidence, and with symmetry in the plane of incidence, that is the $yz$ plane, we follow Refs.~\cite{Menzel:2010-53811:PRA, Zhujun2021ContinuousConversion} and perform the symmetry transformation of the Jones matrix as
\begin{equation} \label{eq:Jsymm}
    J(-\theta) = M_{yz}J(\theta)M_{yz}^{-1} = \begin{pmatrix}
          t_{HH} & -t_{HV} \\ -t_{VH} & t_{VV} 

    \end{pmatrix} ,
\end{equation}
where 
%
%
%
\begin{equation}
M_{yz} = \begin{pmatrix}
        1 & 0 \\ 0 & -1
\end{pmatrix}
\end{equation}
is the reflection matrix.
\textcolor{black}{The coordinate system also transforms when applying the reflection matrix such that $(x,y)\longrightarrow(x,-y)$, or the right-handed coordinate system becomes the left-handed coordinate system and vice-versa. }The equation implies that the Jones matrices for $\pm\theta$ incidence are related through the opposite signs in their anti-diagonal components, or equivalently, their cross-polarizations. We note that Eq.~(\ref{eq:J_unitary}) does not apply in our case, since we consider non-unitary transformations.
On the other hand, if circular dichroism is engineered for $J(\theta)$, then $J(-\theta)$ will also possess circular dichroism but of the opposite handedness. Specifically, after transforming to the circular basis~\cite{Menzel:2010-53811:PRA}, we see that the Jones matrix elements for the two circular polarizations (denoted by $R$ and $L$) are simply exchanged:
\begin{equation}
J^{\rm (circ)}(-\theta) = M_{yz}^{\rm (circ)}J^{\rm (circ)}(\theta)\left(M_{yz}^{\rm (circ)}\right)^{-1} = 
  \begin{pmatrix}
        t_{LL} & t_{LR} \\ t_{RL} & t_{RR}
  \end{pmatrix} ,
\end{equation}
where
\begin{equation}
J^{\rm (circ)}(\theta) = 
\begin{pmatrix}
        t_{RR} & t_{RL} \\ t_{LR} & t_{LL}
  \end{pmatrix} , \quad 
  M_{yz}^{\rm (circ)} = \begin{pmatrix}
        0 & 1 \\ 1 & 0
\end{pmatrix} .
\end{equation}
Furthermore, at normal incidence $\theta = 0$, the Jones matrix must be diagonal,
\begin{equation}
    J(\SI{0}{\degree}) = \begin{pmatrix}
        t_{HH} & 0 \\ 0 & t_{VV}
    \end{pmatrix} ,
\end{equation}
because the symmetry ensures that there are no cross-polarization terms in the linear basis~\cite{Menzel:2010-53811:PRA}. 
Thus, at normal incidence, the eigen-polarizations of the metasurface are linear horizontal and vertical, and they 
remain linear after passing through the metasurface. 
We then perform the singular value decomposition (SVD), which factorizes $J = U\Sigma W^\ast$ where $U$ and $W^\ast$ are unitary by construction and $\Sigma$ is diagonal. The columns of $U$ are eigenvectors of $JJ^\ast$, and the columns of $W$ are the eigenvectors of $J^\ast J$. We refer to the columns of $U$ and $W$ as the left and right singular vectors of $J$ respectively. At normal incidence, we already have $J$ being diagonal, which means that $J$ is normal. Then the product matrices are diagonal and satisfy $J^\ast J = J J^\ast$. Therefore, the columns of $U$ and $W^\ast$ share the same linear vector spaces, which then imply that the left and right singular polarizations are linear. We can tune the degree of linear dichroism from 0 when $\abs{t_{HH}} = \abs{t_{VV}}$, to 1 when $\abs{t_{HH}} \gg \abs{t_{VV}}$ or $\abs{t_{VV}} \gg \abs{t_{HH}}$.

Thus, we propose that it is theoretically possible to design for continuous dichroism by tilting the metasurface, or varying the angle of incidence. The dichroism will generally be elliptical at $\theta\neq0$ and linear at $\theta = 0$. We engineer for this chirality of the system at off-normal incidence and linear dichroism at normal incidence with topology optimization. These metasurfaces may find applications in polarization interferometry or cavity resonance systems. 

To design our metasurface with such a property, we alter the figure of merit to be 
\begin{equation}
     {\rm FOM} = \qty(\abs{J_{-\theta}\ket{RCP}}^2 - \abs{J_{-\theta}\ket{LCP}}^2) \qty(\abs{J_{0}\ket{V}}^2 - \abs{J_{0}\ket{H}}^2).
\end{equation}
and impose mirror symmetry in our device. This symmetry allows us to guarantee that the transmission for circular polarizations in a range of  $+\theta$ incidences will be identical for opposite circular polarizations at $-\theta$ incidences. Thus only one term for $-\theta$ is necessary in the FOM, and we favor transmission of $\ket{RCP}$ at $-\SI{45}{\degree}$ incidence. At normal incidence, 
for our demonstration, we choose the target polarization to be $\ket{V}$. We follow the same optimization principles as outlined previously.  

\begin{figure}[!t]
    \centering
    \includegraphics[width = 0.95\linewidth]{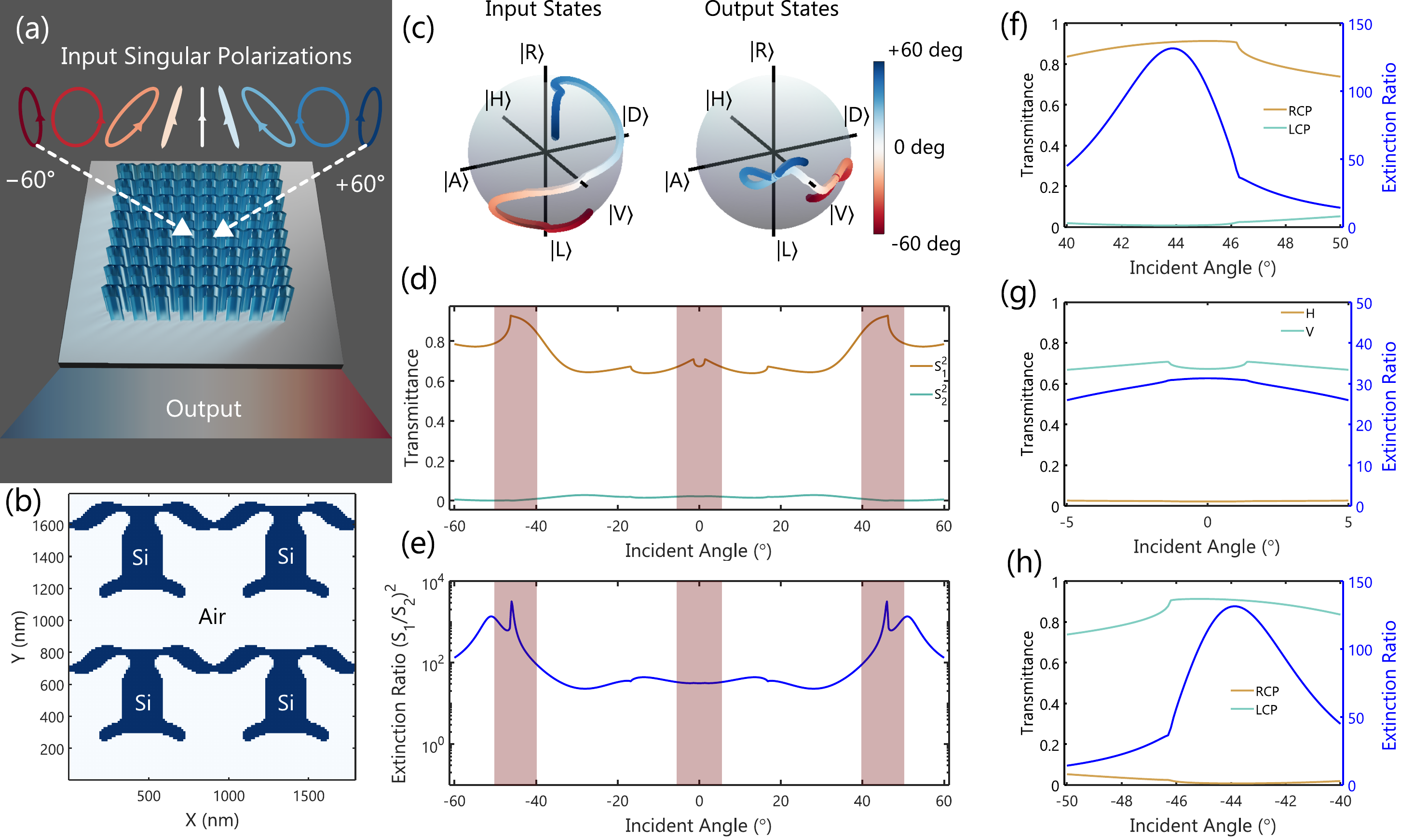}
    \caption{(a) Schematic of continuous angle tunable dichroic metasurface. The input singular polarizations are shown as the incident angle changes from $-\SI{60}{\degree}$ to $+\SI{60}{\degree}$. (b) Optimized metasurface with gratings of silicon surrounded by air. (c) Transformation of input to output singular polarizatio states on the Poincare sphere from $\SIrange{-60}{+60}{\degree}$ incidence angle on the metasurface. (d) Singular values of the the Jones matrix, and (e) corresponding extinction ratio over a range of incidence angles. (f-h) Zoomed in transmission efficiencies for different input polarizations and their respective extinction values for the shaded regions marked in (d,e).  }
    \label{fig:circ_pol_sim}
\end{figure}

The resulting device exhibits continuous elliptical dichroism for a wide range of incidence angles (Fig.~\ref{fig:circ_pol_sim}(a)). Our optimized periodic unit cell has a highly unintuitive connected grating form (Fig.~\ref{fig:circ_pol_sim}(b)). We plot the input singular polarizations, defined as the polarization state that possesses the largest transmission, on the Poincare sphere for incident angles of $-\SI{60}{\degree}$ to $+\SI{60}{\degree}$ (Fig.~\ref{fig:circ_pol_sim}(c)). The path that is traced on the Poincare sphere as the angle of incidence varies passes near the vertical poles of the sphere. This indicates that the singular polarizations are almost circular in nature at $\SI{\pm45}{\degree}$ incidence. As the incidence nears normal, the singular polarizations coalesce into linear $\ket{H}$. For in-between incident angles, the singular polarization states continuously evolve, from elliptical states to circular states and to the linear state at specific angles. The output singular polarizations can take any elliptical state but are coincidentally near the $\ket{V}$ pole of the Poincare sphere.  

\begin{figure}[!t]
    \centering
    \includegraphics[width = 0.95\linewidth]{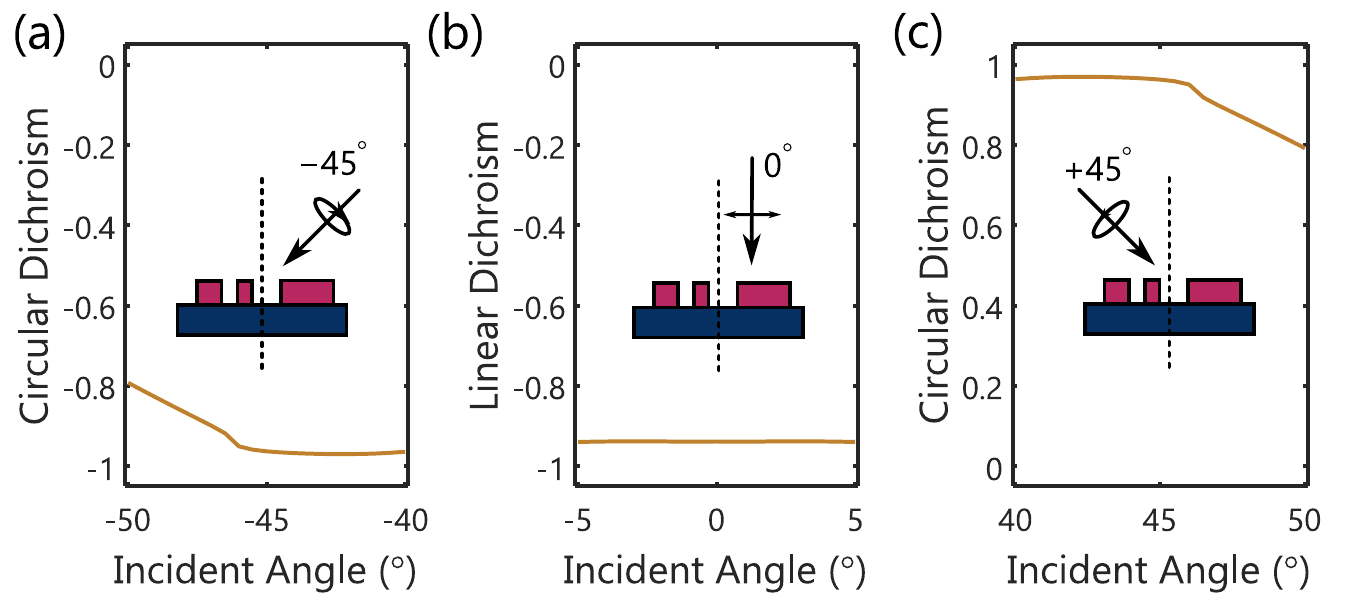}
    \caption{Circular dichroism (a,c) and linear (b) plots for the metasurface design. They are plotted in the intended angular range of operation, $\pm\SI{45}{\degree}$ for circular and \SI{0}{\degree}. }
    \label{fig:CD}
\end{figure}

Our metasurface maintains high transmission across the entire angular range for a continuously varying input polarization (Fig.~\ref{fig:circ_pol_sim}(d)). The symmetry of the  metasurface ensures that the transmittance profile of the singular states is identical for $-\theta$ to $+\theta$ mirrored about normal incidence. The singular value ratio indicates the degree of polarization dichroism, and we see that this value remains large for all incident angles (Fig.~\ref{fig:circ_pol_sim}(e)). Therefore, the metasurface can be considered to have continuous and large dichroism for all simulated incident angles. We also show the transmission of specific polarization states at $-\SI{45}{\degree}, \SI{0}{\degree}, +\SI{45}{\degree}$ incidence in Figs.~\ref{fig:circ_pol_sim}(f-h). For $-\SI{45}{\degree}$ incidence, the metasurface maximally transmits $\ket{RCP}$ with an extinction ratio over 100. While for $+\SI{45}{\degree}$ incidence, the metasurface maximally transmits the orthogonal $\ket{LCP}$ with the same extinction ratio. The transmission of $\ket{V}$ at normal incidence is approximately \SI{70}{\percent} with extinction ratio peaking at 30.

We may define another quantity that characterises the degree of polarization. Circular dichroism ($CD$) and linear dichroism ($LD$) are defined as 
\begin{equation}
    CD = \frac{T_{LL} - T_{RR}}{T_{LL} + T_{RR}}, \quad LD = \frac{T_{HH} - T_{VV}}{T_{HH} + T_{VV}} ,
\end{equation}
with $T$ denoting intensity transmission of 
circular ($L$ and $R$) and linear ($H$ and $V$) polarization states.
The dichroism values range between -1 and 1, with values of 0 indicating no dichroism. Our metasurface exhibit $CD = \pm0.95$ at $\SI{\pm 45}{\degree}$ respectively (Figs.~\ref{fig:CD}(a,c)), indicating strong preferential transmission of a particular handedness of polarization. Furthermore, $LD = -0.93$ at normal incidence, again indicating strong preferential transmission of the $\ket{V}$ state (Fig.~\ref{fig:CD}(b)). There is changing dichroism for intermediate angles of incidence, according to 
Fig.~\ref{fig:circ_pol_sim}(e), 
for varying complex polarization states. Future work may investigate the possibility to tailor the paths that may be traced on the Poincare sphere as the incident angle is tuned, and the corresponding strength of dichroism.

\section{Conclusion}

In this work, we have demonstrated metasurfaces that exhibit an angle-selective polarization response. Our inverse-design scheme allows for precise tailoring of the incidence angles at which differing polarization behaviors materialize. We employed the versatile inverse design scheme to enable flexible polarization control in combination with different oblique incidences. We present novel designes of optimized metasurfaces that exhibit dichroism at only select angles of incidence or over a wide range of angles. We then demonstrate experimentally the optical polarization performance with our manufactured metasurfaces, closely matching numerical modeling. This opens a path in understanding the possibilities to extend to optimizing metasurfaces for non-local wavefront shaping or optical processing.      
  
\section{Acknowledgements}
This work was supported by the Australian Research Council under the Centre of Excellence for Transformative Meta-Optical Systems (TMOS) (CE200100010) and a research grant (NI210100072). This work was performed in part at the Melbourne Centre for Nanofabrication (MCN) in the Victorian Node of the Australian National Fabrication Facility (ANFF). 

\bibliography{references,ref2}

\end{document}